\documentclass[11pt,a4paper]{article}

\usepackage{amsmath}
\usepackage{vmargin}
\usepackage{amsfonts}
\usepackage{float}
\usepackage[parfill]{parskip}
\usepackage[labelfont=bf]{caption}
\usepackage{graphicx}
\usepackage[round,comma]{natbib}
\usepackage{bbm}

\newcommand{\be}{\begin{equation}}
\newcommand{\ee}{\end{equation}}
\newcommand{\ba}{\begin{equation} \begin{aligned}}
\newcommand{\ea}{\end{aligned} \end{equation}}
\newcommand{\ddt}[1]{\frac{\mathrm{d}#1}{\mathrm{d}t}}
\newcommand{\pdt}[2]{\frac{\partial #1}{\partial #2}}

\newcommand{\dint}[1]{\mathrm{d}#1}
\newcommand{\myvec}[1]{ \mathbf{#1} }
\newcommand{\mymat}[1]{ \mathbf{#1} }

\newcommand{\mygvec}[1]{ \boldsymbol{#1} }
\newcommand{\unit}{\mathbbm{1}}

\newcommand{\TR}[1]{#1^{\!\top}}

\title{\bf Hessian corrections to Hybrid Monte Carlo}
\author{Thomas House\\
\textit{School of Mathematics, University of Manchester, Manchester, M13 9PL, UK.}}
\date{}

\begin{document}

\maketitle

\begin{abstract}
	\noindent{}A method for the introduction of second-order derivatives of the
	log likelihood into HMC algorithms is introduced, which does not require the
	Hessian to be evaluated at each leapfrog step but only at the start and end
	of trajectories.
\end{abstract}

\section{Introduction}

Markov chain Monte Carlo (MCMC) is a highly influential computationally
intensive method for performing Bayesian inference, with a large variety of
applications \citep{Brooks:2011}.  While earlier MCMC algorithms made use of
random walks in parameter space \citep{Gilks:1995}, as highlighted in a recent
review by \citet{Green:2015}, the use of derivatives can lead to improved
algorithms.

One derivative-based approach is HMC, standing for either Hybrid
\citep{Duane:1987} or Hamiltonian \citep{Neal:2011} Monte Carlo, which requires
first derivatives of the log likelihood to be available.  More recently, second
derivatives have been included via the use of geometric approaches
\citep{Girolami:2011,Betancourt:2013}. This working paper introduces a
different route to inclusion of second-order derivatives through truncated
Taylor expansion of the log-likelihood, after which Hamilton's equations can be
solved exactly without further approximation. This algorithm is called HHMC
(for Hessian-corrected HMC) and is able to sample accurately from distributions
with different scales for each parameter, which is challenging for standard
HMC.

\section{A Hessian HMC algorithm}

\subsection{Local solution of Hamilton's equations}

The idea behind HMC is to propose a new set of parameters $\mygvec{\theta}^*$
starting from $\mygvec{\theta}^n$ by making use of additional `momentum'
variables $\myvec{p}$. We will consider the general case where $\myvec{p}^n
\sim \mathcal{N}(\myvec{q}, \mymat{Q})$ in general, although in standard HMC,
$\myvec{q} = \myvec{0}$ and $\mymat{Q} = \unit$.

Supposing our aim is to sample according to $\pi$, and we let $l =
\mathrm{ln}(\pi)$, then the basis for HMC algorithm design is Hamilton's equations.
These make use of the Hamiltonian
\be
\mathcal{H} = \frac12 \myvec{p}^2 - l \text{ ,}
\ee
and take the form
\be
\ddt{\mygvec{\theta}} = \pdt{\mathcal{H}}{\myvec{p}} = \myvec{p} \text{ ,}
\qquad
\ddt{\myvec{p}} = - \pdt{\mathcal{H}}{\mygvec{\theta}} = \boldsymbol{\partial}l \text{ .}
\label{Heqs}
\ee
In general, these cannot be solved analytically and so proposals are made on the
basis of a numerical approximation, for example the leapfrog method, in which $L$
steps of length $\varepsilon$ are 
\ba
\mygvec{\theta}_0 & = \mygvec{\theta}^n \text{ ,}\\
\myvec{p}_{1/2} & = \myvec{p}^n + \frac{\varepsilon}{2} \mygvec{\partial}l(\mygvec{\theta}^n)
\text{ ,}\\
\mygvec{\theta}_i & = \mygvec{\theta}_{i-1} + \varepsilon \myvec{p}_{i-(1/2)} \text{ ,} \quad
i \in \{ 1,  \ldots L \} \text{ ,} \\
\myvec{p}_{i+(1/2)} & = \myvec{p}_{i-(1/2)} + \varepsilon \mygvec{\partial}l(\mygvec{\theta}_i)
\quad i \in \{ 1,  \ldots L-1 \} \text{ ,} \\
\mygvec{\theta}^* & = \mygvec{\theta}_L \text{ ,}\\
\myvec{p}^* & = \myvec{p}_{L-(1/2)} + \frac{\varepsilon}{2} \mygvec{\partial}l(\mygvec{\theta}^*)
\text{ .}
\label{leapfrog}
\ea
This algorithm, together with the initial random choice of $\myvec{p}^n$,
defines a marginal proposal density $\rho(\mygvec{\theta}^*|
\mygvec{\theta}^n)$. This will be close to the solution of~\eqref{Heqs} over a
time period $\delta := \varepsilon L$ for large $L$ and small $\varepsilon$. 

Now suppose that we approximate $l = \mathrm{ln}(\pi)$ in the neighbourhood of
some value $\mygvec{\theta}^n$ through Taylor expansion
\be
l(\mygvec{\theta}^n + \myvec{x}) \approx l(\mygvec{\theta}^n) +
\TR{\myvec{v}}\myvec{x} + \frac{1}{2} \TR{\myvec{x}}\mymat{H} \myvec{x} \text{ ,}
\label{taylor}
\ee
where
\be
v_i := \left. \frac{\partial l}{\partial \theta^i} \right|_{\mygvec{\theta}^n} \text{ ,} \quad
H_{ij} := \left. \frac{\partial^2 l}{\partial \theta^i\partial \theta^j}
\right|_{\mygvec{\theta}^n} \text{ ,} \quad
\myvec{v} := (v_i) \text{ ,} \quad
\mymat{H} := (H_{ij}) \text{ .}
\ee

Then we can approximate Hamilton's equations in the region of $\mygvec{\theta}^n$ through
the linear SDE
\be
\dint{\myvec{x}} = \left(\mymat{A}\myvec{x} + \myvec{b} \right)
\dint{t} \text{ ,}
\ee
where
\be
\myvec{x} = \begin{pmatrix} \mygvec{\theta} \\ \myvec{p}
\end{pmatrix} \text{ ,} \quad
\mymat{A}= \begin{pmatrix} \mymat{0} & \unit \\ \mymat{H} & \mymat{0} 
\end{pmatrix} \text{ ,} \quad
\myvec{b}= \begin{pmatrix} \myvec{0} \\ \myvec{v} 
\end{pmatrix} \text{ .} \quad
\ee
Note that this SDE does not have a white-noise term, but is nevertheless a
special case of the results of \citet{Archambeau:2007}, and therefore has
Gaussian solution with mean $\myvec{m}$ and covariance matrix $\mymat{S}$
obeying
\be
\ddt{\myvec{m}} = \mymat{A}\myvec{m} + \myvec{b}
\text{ ,} \qquad
\ddt{\mymat{S}} = {\mymat{A}}\mymat{S}+\mymat{S}\TR{\mymat{A}}
\text{ ,} \qquad
\myvec{m}(0) = \begin{pmatrix} \mygvec{\theta}^n \\ \myvec{q}
\end{pmatrix}\text{ ,} \qquad
\mymat{S}(0) = \begin{pmatrix} \mymat{0} & \mymat{0}\\ \mymat{0} &
\mymat{Q}\end{pmatrix} \text{ .} \label{mSODE}
\ee
Solving these ODEs over the interval $[0, \delta]$ gives, after some analytic
work,
\ba
\mathrm{Mean}(\mygvec{\theta}^*(\delta)) & =
\left( \mathrm{cosh}\left(\mymat{H}^{1/2}\delta\right) - \unit \right)
\mymat{H}^{-1} \myvec{q}
\text{ ,}\\
\mathrm{Cov}(\mygvec{\theta}^*(\delta)) & =
\mymat{H}^{-1/2} \mathrm{sinh}\left(\mymat{H}^{1/2}\delta\right)
 \mymat{Q}
\mymat{H}^{-1/2} \mathrm{sinh}\left(\mymat{H}^{1/2}\delta\right)
\text{ .}
\ea
If we then impose that this should correspond to a situaion where the marginal
proposal $\rho(\mygvec{\theta}^*| \mygvec{\theta}^n) = \pi(\mygvec{\theta}^*)$
under the approximation~\eqref{taylor}, then we obtain a solution for the
proposal distribution after additional analytic work:
\be
\myvec{q}(\mygvec{\theta}^n) =
\mathrm{cosh}\left(\mymat{H}^{1/2}\delta\right) \mymat{H}^{-1/2} 
\left(\mathrm{sinh}\left(\mymat{H}^{1/2}\delta\right)\right)^{-1}\mygvec{\partial}l
\text{ ;}\qquad
\mymat{Q}(\mygvec{\theta}^n) = -
\left(\mathrm{sinh}\left(\mymat{H}^{1/2}\delta\right)\right)^{-2}
\text{ .}\label{sol}
\ee
In terms of numerical computation of~\eqref{sol}, despite the seeming
complexity of the matrix functions, the general approach of \citet{Davies:2003}
is applicable.  In particular, despite the appearance of fractional powers in
the compact expression~\eqref{sol}, all of the functions involved have only
integer matrix powers in their Taylor series and so is expressible as matrix
polynomials.

Note that we can no longer write the accept-reject probabilities in this
algorithm in terms of the Hamiltonian, however since the leapfrog
integrator~\eqref{leapfrog} is reversible it is still possible to calculate
these probabilities. In particular, if we start with $\mygvec{\theta}^n,
\myvec{p}^n$ and propose $\mygvec{\theta}^*, \myvec{p}^*$,
\be
\frac{\rho(\mygvec{\theta}^n, \myvec{p}^n | \mygvec{\theta}^*, \myvec{p}^*)}%
{\rho(\mygvec{\theta}^*, \myvec{p}^* | \mygvec{\theta}^n, \myvec{p}^n)} =
\frac{\mathcal{N}(-\myvec{p}^* | \myvec{q}(\mygvec{\theta}^*),\mymat{Q}(\mygvec{\theta}^*))}%
{\mathcal{N}(\myvec{p}^n | \myvec{q}(\mygvec{\theta}^n),\mymat{Q}(\mygvec{\theta}^n))}
\label{ar}
\ee
The MCMC algorighm based on~\eqref{sol} and~\eqref{ar} is called HHMC for
Hessian-corrected HMC. This algorithm does not require the Hessian $\mymat{H}$
at each leapfrog step, just at the start and end of trajectories, and does not
involve third-order derivatives $\partial_i \mymat{H}$ making it much less
computationally costly than Riemannian approaches
\citep{Girolami:2011,Betancourt:2013}.

\section{A target distribution with heterogeneous scales}

Following \citet{Neal:2012}, consider the following multivariate Gaussian
target density:
\be
\pi(\mygvec{\theta}) = \mathcal{N}(\mygvec{\theta}
| \myvec{0}; \mymat{V})\text{,} \label{nealpi}
\ee
where $\mymat{V}$ is a diagonal matrix with entries equal to the squares of:
110; 100; twenty-six evenly spaced standard deviations between 16 and 8; 1.1;
and 1.0.  Neal suggests this distribution as a diagnostic for HMC because the
variable scales associated with each parameter create difficulties for the
algorithm.

\begin{figure}[H]
\centering
\includegraphics[width=0.98\textwidth]{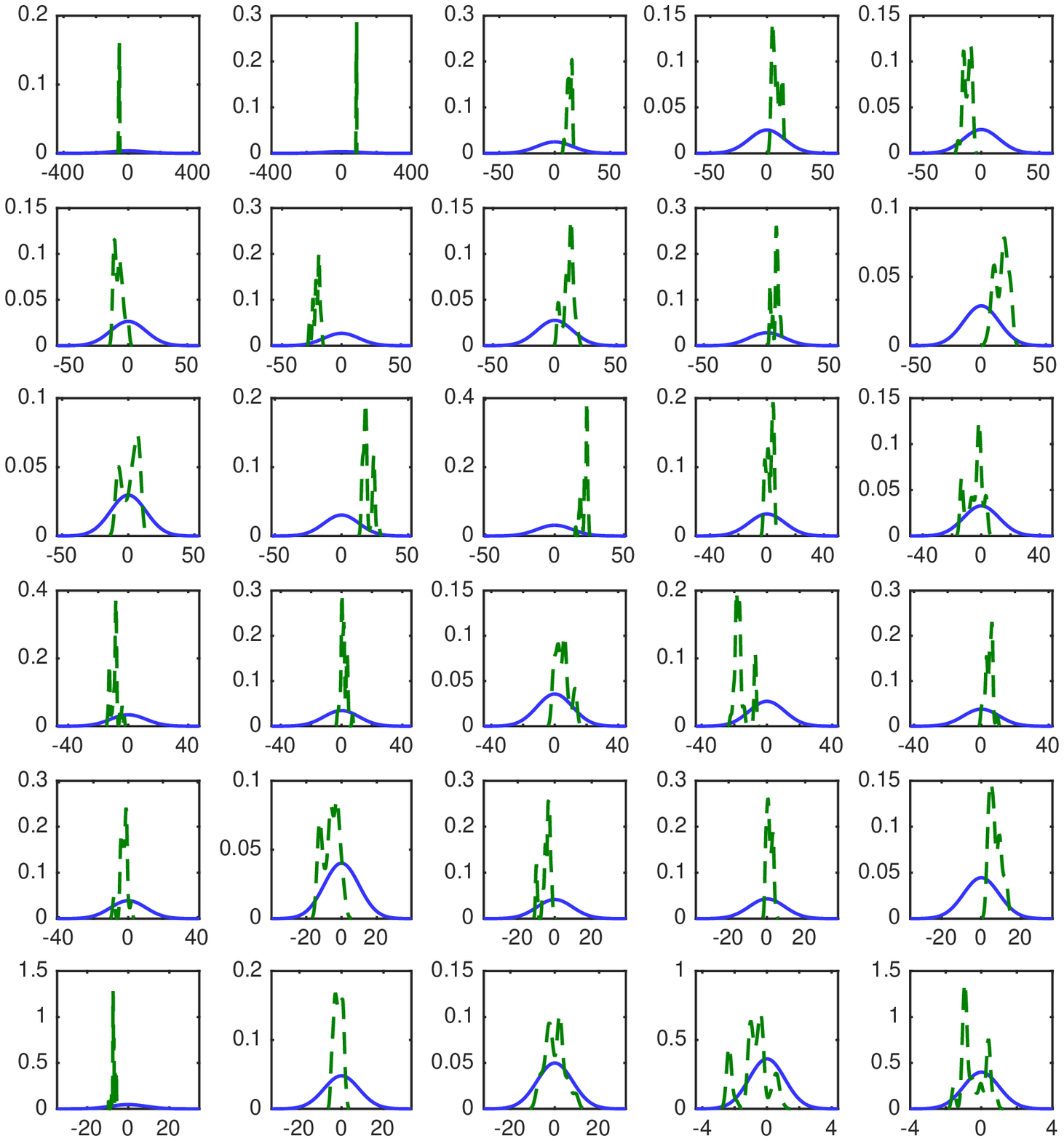}
\caption{Given a multivariate normal target density (blue solid lines) with
	variable scales, HMC (green dashed lines) does poorly. \label{fig:mvn1}}
\end{figure}

\begin{figure}[H]
\centering
\includegraphics[width=0.98\textwidth]{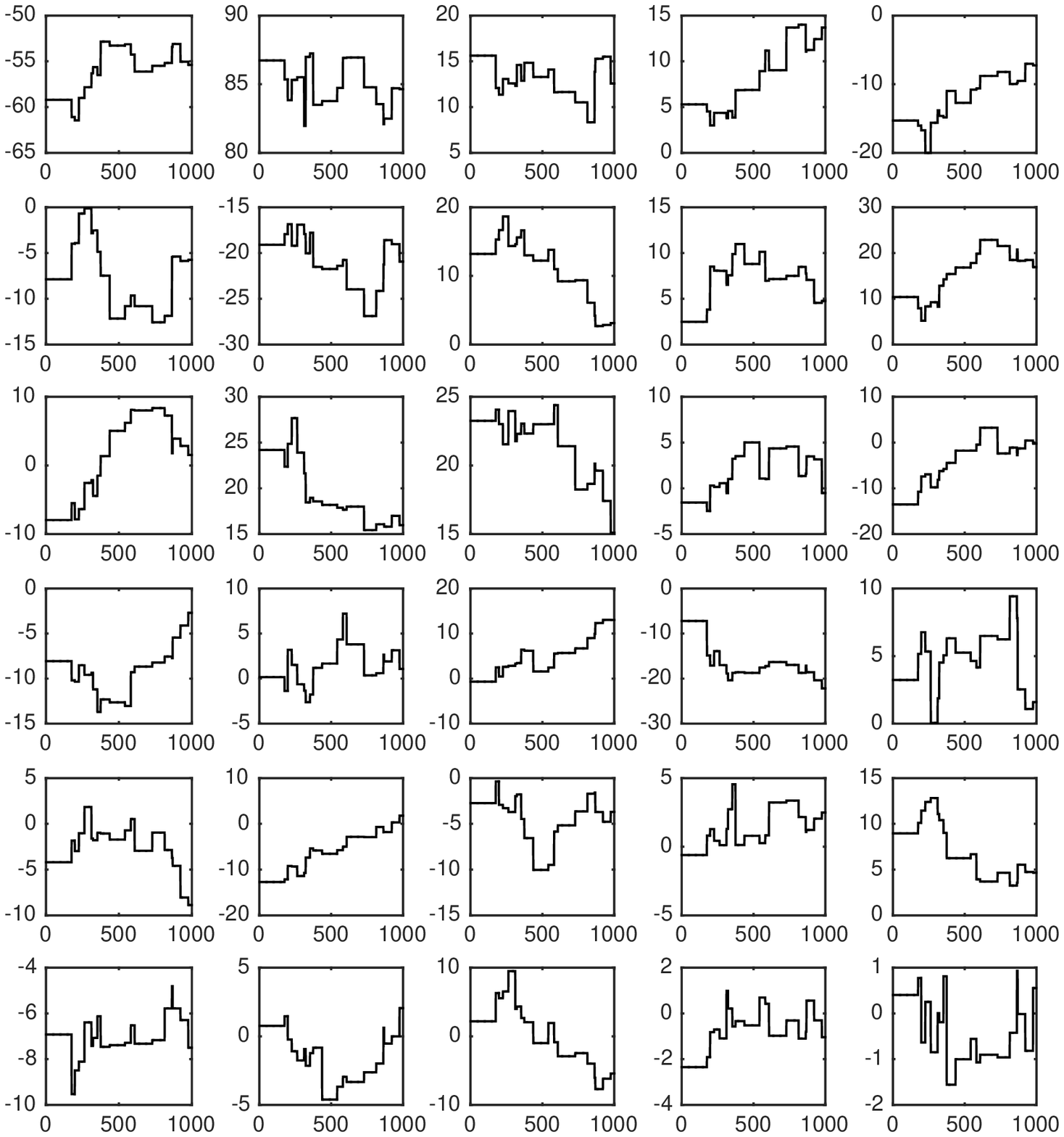}
\caption{HMC trace plots. \label{fig:mvnt1}}
\end{figure}

\begin{figure}[H]
\centering
\includegraphics[width=0.98\textwidth]{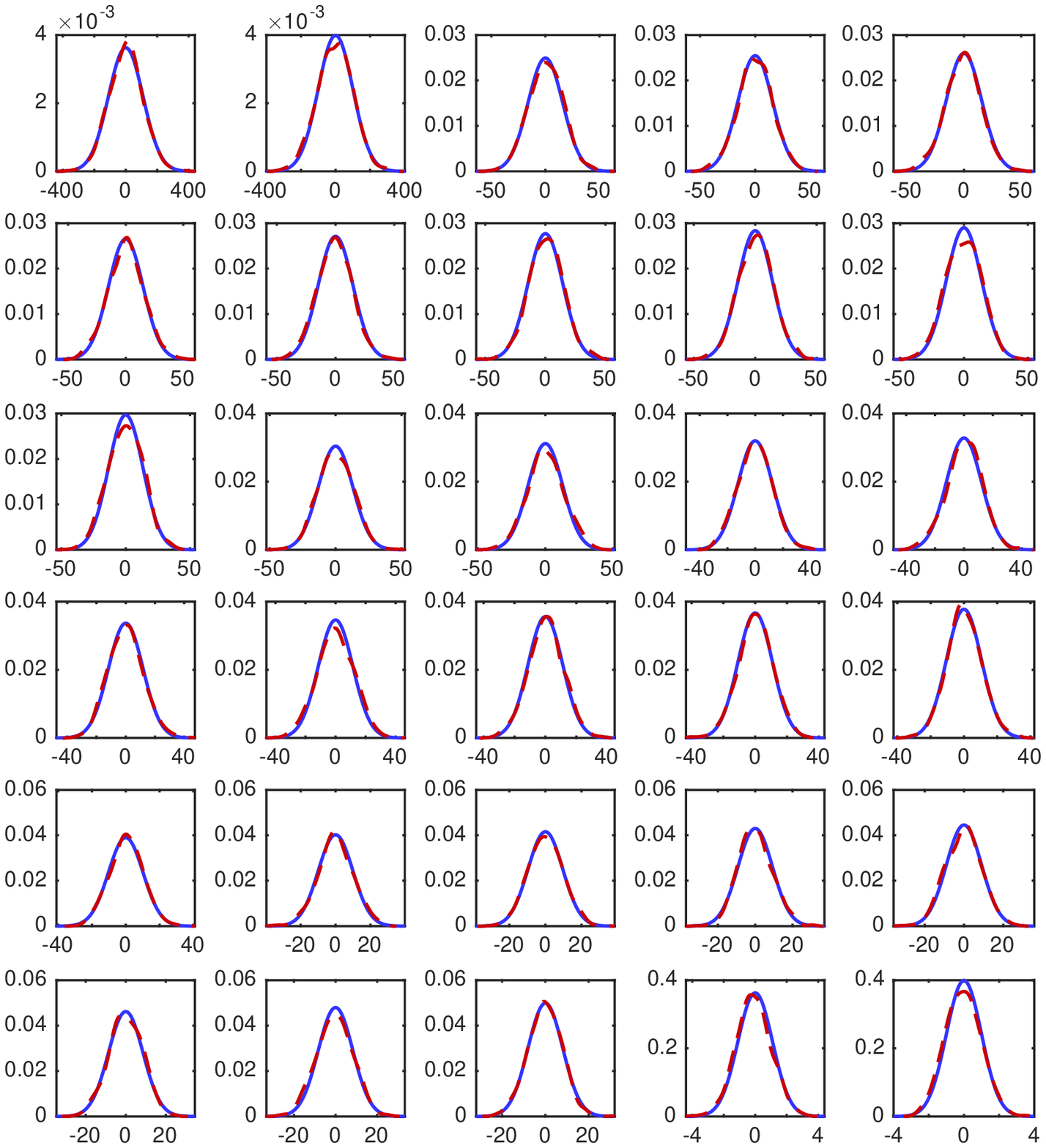}
\caption{Given a multivariate normal target density (blue solid lines) with
	variable scales, HHMC (red dashed lines) does well.  \label{fig:mvn2}}
\end{figure}

\begin{figure}[H]
\centering
\includegraphics[width=0.98\textwidth]{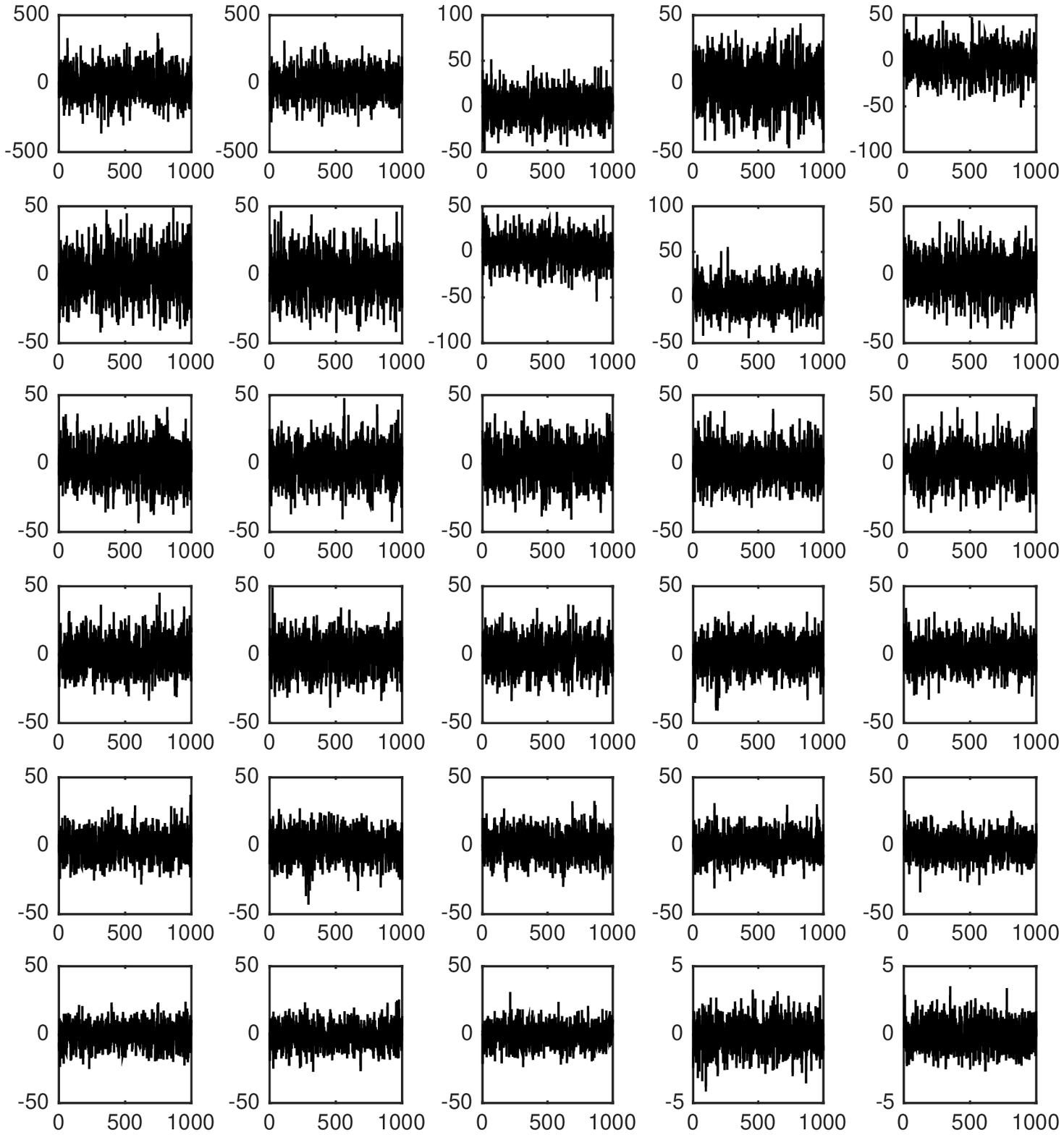}
\caption{HHMC trace plots.  \label{fig:mvnt2}}
\end{figure}

Running MCMC chains of length $10^3$ using $L= 10$ leapfrog steps each with a
step size of $\epsilon = 0.2$ for HMC and HHMC gives the results shown in
Figures~\ref{fig:mvn1}, \ref{fig:mvnt1}, \ref{fig:mvn2} and~\ref{fig:mvnt2}.
The results show that HMC is out-performed by HHMC.  

\section*{Acknowledgements}

Work supported by the UK Engineering and Physical Sciences Research Council.

\clearpage

\clearpage

\clearpage

\clearpage

\end{document}